\documentclass[letterpaper, paper,11pt]{AAS}		

\usepackage{bm}
\usepackage{amsmath}
\usepackage{subfigure}
\usepackage[colorlinks=true, pdfstartview=FitV, linkcolor=black, citecolor= black, urlcolor= black]{hyperref}
\usepackage{overcite}
\usepackage{footnpag}			      	
\usepackage{hyperref}
\usepackage{footmisc}
\usepackage{svg}
\usepackage{tabularx}
\usepackage{graphicx}

\PaperNumber{2024-406}

\begin{document}

\newcommand{\affilA}{\footnotemark[1]} 
\newcommand{\affilB}{\footnotemark[2]} 
\newcommand{\affilC}{\footnotemark[3]} 
\newcommand{\affilD}{\footnotemark[4]} 
\newcommand{\affilE}{\footnotemark[5]} 
\newcommand{\affilF}{\footnotemark[6]} 
\title{HERA ORBIT DETERMINATION COVARIANCE ANALYSIS OF DIDYMOS BINARY ASTEROID WITH THE FULL TWO-BODY PROBLEM}
\author{Edoardo Gramigna\affilA\,\affilB,
Riccardo Lasagni Manghi\affilA,
Marco Zannoni\affilA\,\affilB,
Paolo Tortora\affilA\,\affilB,
Michalis Gaitanas\affilC,
Ioannis Gkolias\affilC,
Kleomenis Tsiganis\affilC,
Ryan S. Park\affilD,
Alex Meyer\affilE,
Petr Pravec\affilF}

\maketitle{} 		
\footnotetext[1]{Department of Industrial Engineering (DIN), University of Bologna, via Fontanelle 40, 47121, Forlì, Italy. edoardo.gramigna@unibo.it}
\footnotetext[2]{Centro Interdipartimentale di Ricerca Industriale Aerospaziale (CIRI AERO) Alma Mater Studiorum – Università di Bologna, Via Baldassarre Carnaccini 12, 47121 Forlì, Italy.}
\footnotetext[3]{Section of Astrophysics, Astronomy and Mechanics, Department of Physics, Aristotle University of Thessaloniki, 54124 Thessaloniki, Greece.}
\footnotetext[4]{NASA Jet Propulsion Laboratory, California Institute of Technology, Pasadena, CA, USA.}
\footnotetext[5]{Smead Department of Aerospace Engineering Sciences, University of Colorado Boulder, Boulder, CO, USA} 
\footnotetext[6]{Astronomical Institute, Academy of Sciences of the Czech Republic, Ond\v{r}ejov, CZ-25165, Czech Republic.} 
\begin{abstract}
Standard planetary radio science analyses may be inadequate for accurately interpreting data from the Hera mission at the Didymos system due to the anticipated challenge posed by the coupling between the translational and rotational motions of the asteroids. A more precise dynamical model, the full two-body problem model, which includes a mutual gravitational potential, may be required to fit the Hera data properly. This work presents a covariance analysis of the Hera radio science experiment using this advanced model. The tested dynamical scenarios are based on recent lightcurve observations indicating that Dimorphos is in an excited non-principal axis spin state that is, on average, tidally locked. Three non-principal axis scenarios are analyzed using a multi-arc local estimation approach instead of the global one adopted in previous investigations involving non-chaotic states. The results show that tumbling or chaotic cases significantly improve the recovery of the bodies' moments of inertia due to the enhanced observability of the system's dynamics from the excited spin state. However, the local estimation approach slightly degrades the accuracy of the mass estimation by a factor of 1.3-3 and Dimorphos' relative orbit by about one order of magnitude. Developing a dedicated constrained multi-arc filter is thus required to improve the estimation accuracy of the Didymos system's dynamical state further. Nevertheless, these results still meet the Hera radio science requirements.

\end{abstract}

\section{Introduction}
Hera represents the European Space Agency's inaugural planetary defense space mission \cite{michel2022hera} and plays a pivotal role in the Asteroid Impact and Deflection Assessment international collaboration \cite{cheng2018aida} alongside NASA's DART mission. DART, last 26th September 2022, conducted the first asteroid deflection experiment on Dimorphos, the small moon of the Didymos binary asteroid system, using the kinetic impactor technique \cite{cheng2023momentum}. The European counterpart, Hera, aims to comprehensively assess and characterize the feasibility of the kinetic impactor technique for an asteroid deflection while conducting an in-depth investigation of the binary asteroid. The Hera investigation includes examining its physical and compositional properties and studying the effect of the impact on the surface and shape of Dimorphos.

In this context, the Hera radio science experiment\cite{gramigna2024hera} will precisely estimate key physical parameters necessary to determine the momentum enhancement resulting from the DART impact. These parameters include the mass, mass distribution, rotational states, relative orbits, and dynamics of Didymos and Dimorphos. The gravity, orbits, and rotational states of Didymos and Dimorphos can be estimated by precisely reconstructing the trajectories of the Hera, Juventas, and Milani spacecraft during the nominal asteroid phase \cite{gramigna2022hera}. The expected results of the standard radio science investigation for Hera are presented in Gramigna et al. (2024) exploiting a covariance analysis.\cite{gramigna2024hera}. 

However, the application of standard planetary radio science analysis, which traditionally involves rotational states treated independently and described by polynomials that are functions of time, may prove insufficient for accurately interpreting the data gathered by the Hera mission. This limitation arises due to the anticipated challenge posed by the coupling between the translational and rotational motions of the asteroids. For these reasons, in previous work, we implemented the full two-body problem model, a higher-accuracy dynamical model characterized by a mutual gravitational potential, and we performed covariance analysis exploiting this model of pre-impact relaxed and post-impact non-chaotic scenarios. \cite{gramigna2024hera}

Following the recent findings published by Pravec et al. (2024)\cite{pravec2024rotational}, their observations indicate that Dimorphos exhibits an excited spin state characterized by non-principal axis (NPA) rotation while being, on average, tidally locked. By comparing these observations with detailed simulations, they determined that a moderate degree of NPA rotation post-DART impact is necessary to match the observed lightcurves. This rotation is restricted to configurations that are typically synchronous or anti-synchronous on average. Consequently, in this work, we perform a covariance analysis of the Hera RSE to examine some of these expected NPA dynamical scenarios. Given the chaotic nature of these dynamical systems, we adopt a local approach for estimating the asteroids' ephemerides and rotational states, as opposed to the global approach used by Gramigna et al. (2024)\cite{gramigna2024hera}.

The following sections present the orbit determination setup, a comprehensive description of the implemented two-body problem model, and its validation. Subsequently, we characterize the dynamical modeling of the Didymos system and the expected NPA scenarios for Dimorphos in the aftermath of DART's impact. Finally, we present the results of this investigation, comparing the NPA covariance analysis results to the pre- and post-impact non-chaotic scenarios discussed in our previous work\cite{gramigna2024hera}.

\section{Orbit Determination Setup}
The orbit determination (OD) setup adopted in this work follows the one described in Gramigna et al. (2024)\cite{gramigna2024hera}. The measurements used in the OD process are the Hera two-way Doppler and range at X-band, obtained from the standard Earth-based communication link, the Inter-Satellite Link (ISL) two-way Doppler and range at S-band between Hera and the CubeSats (Juventas and Milani), and the optical navigation images of Didymos and Dimorphos acquired by Hera's Asteroid Framing Camera (AFC) every 2 hours. For this study, we considered the asteroid nominal phase, composed of the Early Characterization Phase (ECP), the Detailed Characterization Phase (DCP), and the Close Observation Phase (COP).\cite{gramigna2024hera}

Overall, the dynamical and observational models, the measurement schedule, the noise characteristics, and the \textit{a priori} values and covariances of the estimated parameters correspond to the ones reported in Gramigna et al. (2024)\cite{gramigna2024hera} and are not repeated here for brevity. The main differences between the two setups are the asteroids' initial conditions and inertia properties. The former are obtained from the Pravec et al. (2024) lightcurve analysis described below for selected non-principal axis scenarios. \cite{pravec2024rotational} The latter are computed assuming a triaxial ellipsoid shape (derived again from Reference\cite{pravec2024rotational}), according to Equation \ref{eqellipsinertia}. 
\begin{flalign}
	&I_{i_{XX}} = \frac{M_i}{5}\left(b_i^2 + c_i^2 \right), I_{i_{YY}} = \frac{M_i}{5}\left(a_i^2 + c_i^2 \right), I_{i_{ZZ}} = \frac{M_i}{5}\left(a_i^2 + b_i^2 \right)
 \label{eqellipsinertia}
\end{flalign}
where $a_i, b_i, c_i$ are the principal axes of the ellipsoid for body $i$, and $M_i$ is its mass, which is retrieved from the ellipsoid volume assuming a constant density.\cite{pravec2024rotational} 

\section{Full two-body problem model}
\begin{figure}[htb]
	\centering\includegraphics[width=0.6\textwidth]{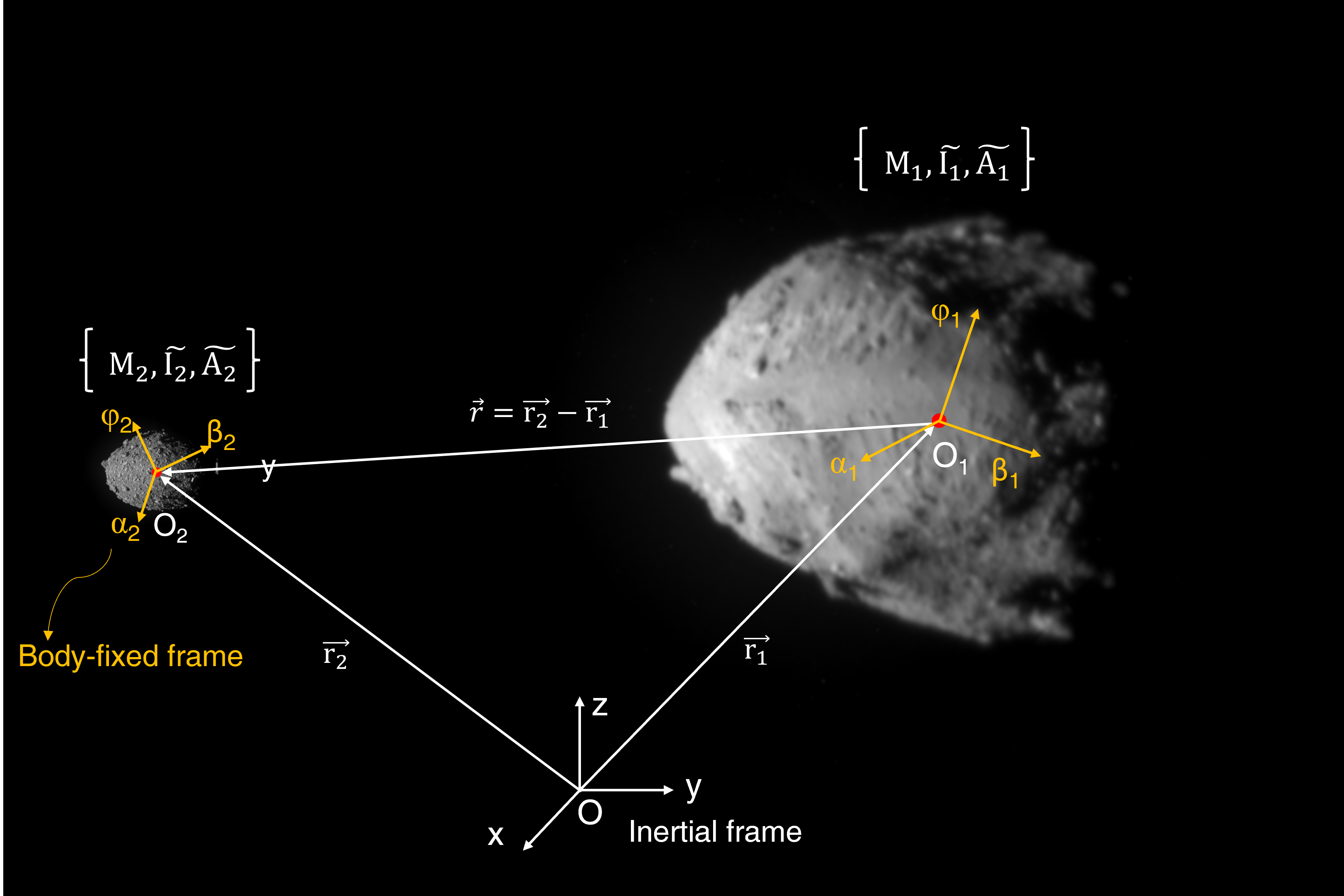}
    \caption{Full two-body problem model. $M_i$, $I_i$, and $A_i$ represent the bodies' masses, inertia moments, and rotation matrices. The image of the Didymos system is from DART DRACO camera mosaics. Image credit: NASA/Johns Hopkins APL, \url{https://dart.jhuapl.edu/Gallery/}.}
	\label{fig:f2bpmodel}
\end{figure}

The Didymos binary system perfectly illustrates the Full Two-Body Problem (F2BP) \cite{maciejewski1995reduction,ashenberg2007mutual}. The F2BP characterizes the dynamic interplay between two mass distributions, leading to interconnected translational and rotational movements driven by the mutual gravitational potential shared between irregular mass distributions. The dynamic evolution of the system is mainly influenced by the shapes, initial positions, and orientations of each component, as depicted in Figure \ref{fig:f2bpmodel}. In this context, standard radio science models (i.e., integration of the translational motion and rotational states defined by polynomials function of time) may not correctly fit the data due to the complex dynamical state. Therefore, simulation codes designed explicitly for the F2BP are required to comprehensively model the intricate dynamics of the system in preparation for Hera data analysis.

Many works in the literature derived the complete formulation of the F2BP; among them, see References \citenum{maciejewski1995reduction,ashenberg2007mutual,scheeres2009stability,hou2017mutual}, where theoretical
details can be found. The F2BP model adopted within this research is a 3D model based on Ashenberg (2007).\cite{ashenberg2007mutual} Its derivation relies on Cartesian coordinates, inertia integrals associated with the principal reference frames of each body, and the relative rotation matrix. 

We implemented the F2BP model in JPL's orbit determination program Mission Analysis, Operations, and Navigation Toolkit Environment (MONTE) \cite{evans2018monte}, which previously lacked support for spin-orbit coupling. This model is then used to perform Hera radio science covariance analysis.

Within this work, the mutual gravitational potential and the mutual gravitational torque of two bodies of arbitrary shape are expanded to the second order, namely:
\begin{flalign}
	U = \underbrace{- \frac{G M_1 M_2}{d}}_\text{$U^{(0)}$} & - \frac{G M_2}{2 d^3} [ I_{1xx} + I_{1yy} + I_{1zz} - 3 ( l_1 ( I_{1xx}l_1 + I_{1xy}m_1 + I_{1xz}n_1 ) +  && \label{eqf2bppotential}\\
		& + m_1 ( I_{1xy}l_1 + I_{1yy}m_1 + I_{1yz}n_1 ) + n_1 ( I_{1xz}l_1 + I_{1yz}m_1 + I_{1zz}n_1 ) ) ] + &&  \notag \\
		& - \frac{G  M_1}{2  d^3}[ I_{2xx} + I_{2yy} + I_{2zz} - 3 ( l_2 ( I_{2xx}l_2 + I_{2xy}m_2 + I_{2xz}n_2 ) + && \notag \\
		& + m_2 ( I_{2xy}l_2 + I_{2yy}m_2 + I_{2yz}n_2 ) + n_2 ( I_{2xz}l_2 + I_{2yz}m_2 + I_{2zz}n_2 ) ) ], && \notag
\end{flalign}
where $G$ is the gravitational constant, $d$ is the bodies' relative distance, $M_{i}$, $I_{i}$, $n_i$-$m_i$-$l_i$ are respectively the mass, inertia tensor, and direction cosines (in body-frame) of the i-th body. In Equation \ref{eqf2bppotential}, we highlighted the zeroth-order potential $U^{(0)}$, corresponding to the point mass contribution. The sum of the remaining terms forms the second-order potential $U^{(2)}$. It should be noted that the first-order potential $U^{(1)}$ is missing from Equation \ref{eqf2bppotential}. This term captures the influence of potential offsets between the body's reference position and its center of mass. We assume the two are precisely aligned in this scenario, so the term $U^{(1)}$ is null.

Assuming to place the origin of the inertial reference frame at the center of mass of the binary
system, the acceleration acting on Dimorphos due to the F2BP is given by Equation \ref{eqaccel1}:
\begin{flalign}
M_2 \overset{\cdot\cdot}{\vec{r_2}} = - \vec{\nabla} U =  - \frac{\partial U}{\partial \vec{r_2}} = - \frac{\partial U}{\partial \vec{r}},
\label{eqaccel1}
\end{flalign}
We can write this expression in explicit form as:
\begin{flalign}
&\overset{\cdot\cdot}{\vec{r_2}} = -\frac{1}{M_2} \Bigg\{ \left( - \frac{U^{(0)}}{d} - \frac{3 U^{(2)}}{d} \right) \left( \frac{\vec{r}}{d} \right) + \left( \frac{3 G M_2 (I_{1xx}l_1 + I_{1xy}m_1 + I_{1xz}n_1)}{d^3} \right) \left( \frac{\vec{a_1}d - \vec{r}l_1}{d^2} \right) + \notag \\
&+\left( \frac{3 G M_2 (I_{1xy}l_1 + I_{1yy}m_1 + I_{1yz}n_1)}{d^3} \right) \left( \frac{\vec{a_2}d - \vec{r}m_1}{d^2} \right) + \left( \frac{3 G M_2 (I_{1xz}l_1 + I_{1yz}m_1 + I_{1zz}n_1)}{d^3} \right) \left( \frac{\vec{a_3}d - \vec{r}n_1}{d^2} \right) + \notag\\
&+\left( \frac{3 G M_1 (I_{2xx}l_2 + I_{2xy}m_2 + I_{2xz}n_1)}{d^3} \right) \left( \frac{\vec{b_1}d - \vec{r}l_2}{d^2} \right) + \left( \frac{3 G M_1 (I_{2xy}l_2 + I_{1yy}m_2 + I_{2yz}n_2)}{d^3} \right) \left( \frac{\vec{b_2}d - \vec{r}m_2}{d^2} \right) + \notag\\
&+\left( \frac{3 G M_1 (I_{2xz}l_2 + I_{2yz}m_2 + I_{2zz}n_2)}{d^3} \right) \left( \frac{\vec{b_3}d - \vec{r}n_2}{d^2} \right) \Bigg\},
\label{eqaccel}
\end{flalign}
where $\vec{r}$ is the asteroids' mutual separation vector, and $\vec{a_i}$, $\vec{b_i}$ are the column vectors of the rotation matrices between the body-fixed frames and the inertial frames of Didymos and Dimorphos, respectively.

The F2BP acceleration derived in Equation \ref{eqaccel} can be directly incorporated into our MONTE base force model for integration. For more details on the asteroids' base force model, see Section 5 in Gramigna et al. (2024) \cite{gramigna2024hera}.

It is worth mentioning that, within the OD process, we numerically integrate only the trajectories of the system's barycenter and Dimorphos. The trajectory of Didymos is uniquely determined by the orbits of the integrated bodies and their masses according to: 
\begin{equation}
	\vec{r}_{C1} = - \frac{GM_2}{GM_1 + G M_2} \vec{r}_{12},
\end{equation}
where $\vec{r}_{AB}$ is a generic position vector from A to B, and the subscripts for the different elements are C for the system's center of mass, 1 for Didymos, and 2 for Dimorphos. Analogous relations hold for velocity and acceleration.

Regarding the rotational motion, let's consider the gravity gradient torques acting on Didymos and Dimorphos expressed in the inertial frame, respectively denoted as $\vec{\tau_{1i}}$ and $\vec{\tau_{2i}}$.
\begin{flalign}
	& \tau_{1i} = - \sum_{n=1}^{3} \vec{a_n} \times \frac{\partial U}{\partial \vec{a_n}} = - \vec{a_1} \times \frac{\partial U}{\partial \vec{a_1}} - \vec{a_2} \times \frac{\partial U}{\partial \vec{a_2}} - \vec{a_3} \times \frac{\partial U}{\partial \vec{a_3}}, \label{eqf2bptorques}\\
	& \tau_{2i} = - \sum_{n=1}^{3} \vec{b_n} \times \frac{\partial U}{\partial \vec{b_n}} = - \vec{b_1} \times \frac{\partial U}{\partial \vec{b_1}} - \vec{b_2} \times \frac{\partial U}{\partial \vec{b_2}} - \vec{b_3} \times \frac{\partial U}{\partial \vec{b_3}}, \notag
\end{flalign}
Here, the symbol $\times$ denotes the cross product between the vectors. In a more explicit form Equation \ref{eqf2bptorques} can be written as:
\begin{flalign}
\tau_{1i} = &- \vec{a_1} \times \left( \frac{3 G M_2 (I_{1xx} l_1 + I_{1xy} m_1 + I_{1xz} n_1)}{d^3} \hat{r} \right) - \vec{a_2} \times \left( \frac{3 G M_2 (I_{1xy} l_1 + I_{1yy} m_1 + I_{1yz} n_1)}{d^3} \hat{r} \right) + \notag \\
&-\vec{a_3} \times \left( \frac{3 G M_2 (I_{1xz} l_1 + I_{1yz} m_1 + I_{1zz} n_1)}{d^3} \hat{r} \right)  ,
\end{flalign}
\begin{flalign}
\tau_{2i} = & - \vec{b_1} \times \left( \frac{3 G M_1 (I_{2xx} l_2 + I_{2xy} m_2 + I_{2xz} n_2)}{d^3} \hat{r} \right) - \vec{b_2} \times \left( \frac{3 G M_1 (I_{2xy} l_2 + I_{2yy} m_2 + I_{2yz} n_2)}{d^3} \hat{r} \right) + \notag \\
&-\vec{b_3} \times \left( \frac{3 G M_1 (I_{2xz} l_2 + I_{2yz} m_2 + I_{2zz} n_2)}{d^3} \hat{r} \right). \notag 
\end{flalign}

Finally, the evaluated F2BP torques are added to the MONTE torque models for the rotational state integration.

The model has been tested and validated against the University of Thessaloniki F2BP propagator, which was previously validated against the General Use Binary Asteroid Simulator (GUBAS)\cite{davis2020doubly} used to estimate DART's momentum enhancement factor in Cheng et al. (2023)\cite{cheng2023momentum}. Furthermore, our MONTE F2BP model underwent a comprehensive comparison with other results presented in the literature, such as the ones produced by Agrusa et al. (2020, 2021)\cite{agrusa2020benchmarking,agrusa2021excited}. In particular, we compare the results reported in Agrusa et al. (2020)\cite{agrusa2020benchmarking} and the ones obtained with our model using the same assumptions and initial conditions.

\begin{figure}[h!]
	\centering
f	\includegraphics[width=0.6\textwidth]{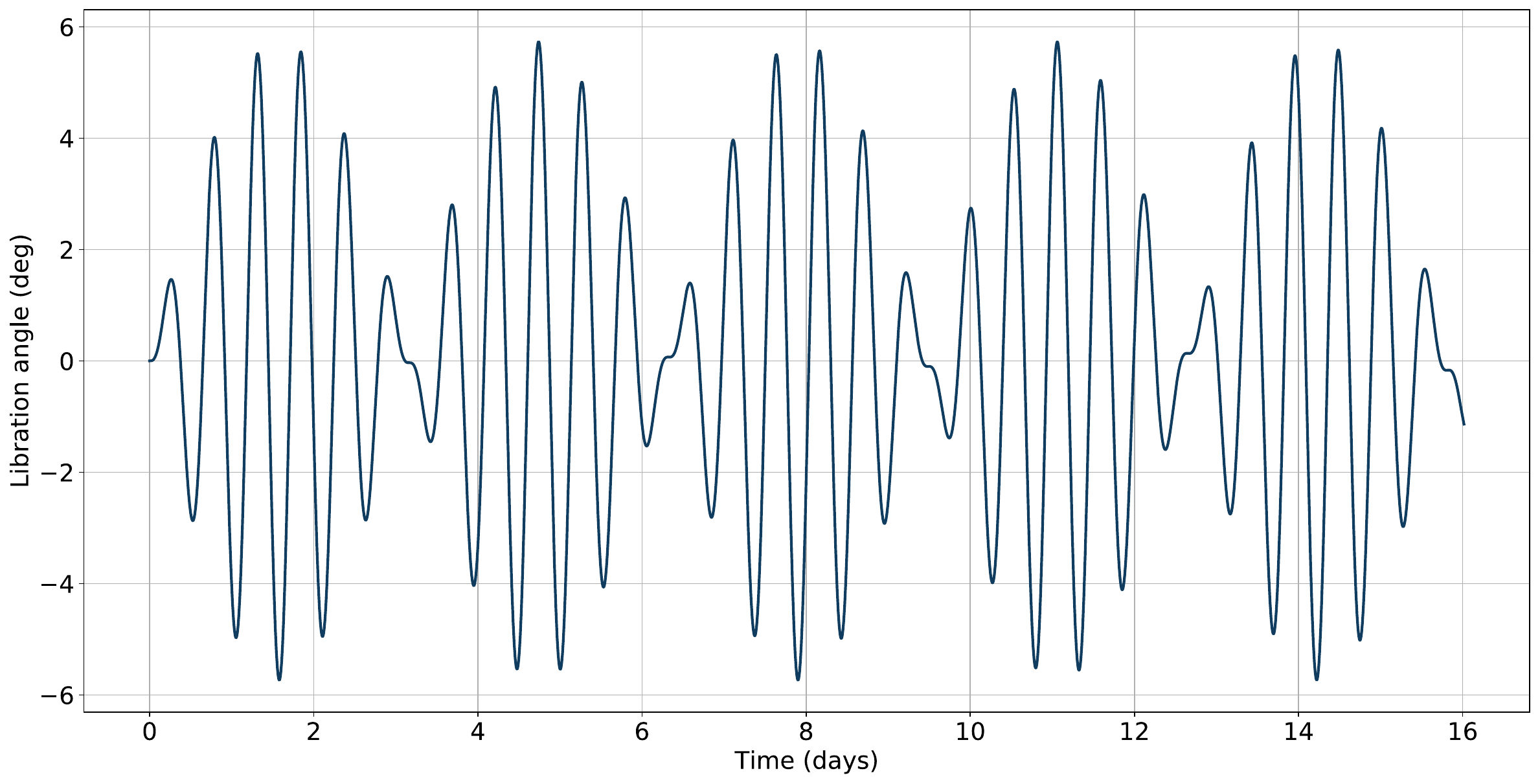}
	\caption{Longitudinal libration of UNIBO MONTE F2BP over 15 days. The libration pattern is consistent over longer propagation times, and the results match the ones presented in Agrusa et al. (2020) \cite{agrusa2020benchmarking}.}
	\label{figf2bpagrusa2020a}
\end{figure}

\begin{figure}[h!]
	\centering
	\includegraphics[width=0.6\textwidth]{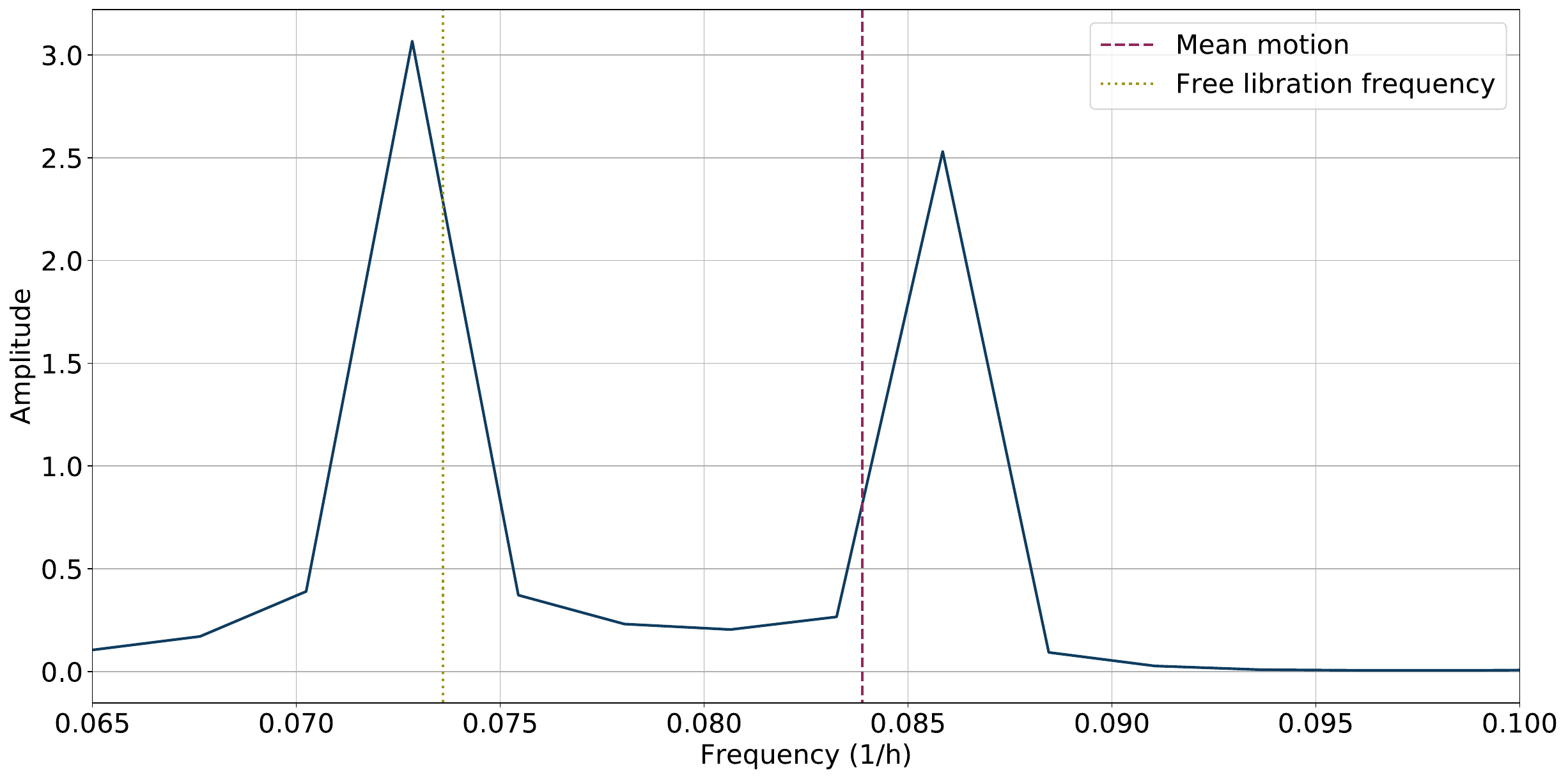}
	\caption{Fourier transform of longitudinal libration from Figure \ref{figf2bpagrusa2020a}.}
	\label{figf2bpagrusa2020b}
\end{figure}

Figure \ref{figf2bpagrusa2020a} illustrates the longitudinal libration of Dimorphos, showing a similar pattern to the one in Agrusa et al. (2020) (see Figure 7a in Reference). The slight difference in the libration amplitude arises from variations in the Didymos shape models used in the two studies, along with our use of a 2nd-order gravity approximation, as opposed to their 4th-order F2BP. Figure \ref{figf2bpagrusa2020b} presents the Fourier transform of the libration in Figure \ref{figf2bpagrusa2020a}, revealing two distinct frequency bands. The first roughly corresponds to the free libration, with a frequency that depends on the body's moments of inertia. The second is the forced libration, occurring at the frequency of the orbital angular velocity oscillation (also referred to as the epicyclic or radial frequency).
In the decoupled spin-orbit motion, the frequency of free libration for a synchronous
satellite on a circular orbit is:
\begin{equation}
	\omega_0 = n\left( 3 \frac{I_{yy} - I_{xx}}{I_{zz}} \right)^{1/2},
	\label{eqfreelib}
\end{equation}
where $n$ is the mean motion, and $I_{xx} < I_{yy} < I_{zz}$ are the three principal moments of inertia of the secondary \cite{murray2000solar}. The forced (relaxed) mode inherently manifests in a synchronous secondary on an eccentric orbit, and its amplitude can be described using Equation \ref{eqforcedampl}: 
\begin{equation}
	w_a = \frac{2e}{1 - \frac{I_{zz}}{3(I_{yy} - I_{xx})}},
	\label{eqforcedampl}
\end{equation}
where $e$ is the eccentricity.

In this scenario, the secondary undergoes a periodic restoring torque with a frequency equal to the mean motion. This torque results from the misalignment of the long axis with the line of centers, a consequence of the orbital angular velocity varying throughout a single orbit. However, the full two-body model introduces additional complexity, as the coupling also influences the librations and their frequencies. For a more in-depth exploration of these two libration modes in coupled and uncoupled scenarios, readers can refer to Reference \cite{naidu2015near}.

About the results presented in Figure \ref{figf2bpagrusa2020b}, it is evident that the two frequencies deviate slightly from the analytical ones, as given by Equation \ref{eqfreelib} for the free libration and the mean motion frequency for the forced libration. This discrepancy can be attributed to the following factors, as noted by Agrusa et al. (2020)\cite{agrusa2020benchmarking}:
\begin{enumerate}
	\item The analytical free libration of Equation \ref{eqfreelib} assumes a spherically symmetric primary on a fixed orbit. In contrast, this simulation incorporates a non-spherically symmetric Didymos with a non-fixed orbit, accounting for the small differences observed.
	\item Concerning the forced libration, the mean motion would precisely match the epicyclic frequency only in case of no apsidal precession (fixed orbit). However, due to the oblate nature of the primary and the resulting orbit, a non-negligible periapse precession occurs, contributing to the observed deviation from the expected mean motion frequency.
\end{enumerate}

\section{Didymos System Dynamical State} 

The fast rotation and top-like configuration of Didymos suggest a potential rubble-pile structure, as indicated by Richardson et al. (2006) \cite{richardson2006binary}. While this prompts further exploration into rubble-pile studies, the initial approach of treating the bodies as rigid for integrating their dynamics is a reasonable first approximation.\cite{agrusa2022dynamical} Given the highly dissipative nature of rubble-pile asteroids, it is also probable that mutual tides have dampened the system to reach an equilibrium state, with Dimorphos in a synchronous rotation and a nearly circular orbit \cite{goldreich2009tidal} before the DART impact. Indeed, the assumption of synchronous rotation is well-supported. Measurements of the mean anomaly drift in the pre-impact orbit suggest the influence of the binary YORP (BYORP) effect, which requires a synchronous secondary.\cite{scheirich2022preimpact,cuk2005effects} Additionally, observed spin rates of secondary bodies in small binary asteroids like Didymos are consistent with synchronous rotations.\cite{pravec2016binary} This observation, along with a low pre-impact eccentricity \cite{scheirich2022preimpact}, supports the pre-impact synchronous assumption. Pravec et al. (2024) \cite{pravec2024rotational} findings further strengthen this argument. First, post-impact lightcurve data aligns with a secondary in near-synchronous rotation; second, high-fidelity simulations beginning with a synchronous state also closely match these observations. Collectively, these points strongly indicate that Dimorphos was in a 1:1 spin-orbit resonance before the DART impact. For these reasons, we followed the approach described by Richardson et al. (2022) \cite{richardson2022predictions} and assumed that the system was relaxed before the impact. By modifying Dimorphos's orbital velocity, the DART impact has promptly modified this relaxed state, inducing changes in both the mutual orbit and spin state of Dimorphos. In this context, Dimorphos' post-impact dynamical state is not yet fully known, and Hera will provide crucial data to characterize the system's dynamics fully.

In a previous work\cite{gramigna2024hera}, we explored a F2BP post-impact non-chaotic case and a pre-impact relaxed scenario through a covariance analysis to assess the sensitivity of the Hera RSE to different dynamical conditions. Furthermore, we compared the two F2BP cases with a \textit{Nominal} scenario, adopting standard radio science models, to study the effect of different dynamical representations within the OD process.

Following the recent results from Pravec et al. (2024)\cite{pravec2024rotational}, the latest lightcurve observations suggest that the Didymos system is in a chaotic or NPA rotational state after the DART impact. Therefore, in this work, we address these newly expected dynamical scenarios to understand how different dynamical regimes may influence the Hera RSE OD process.

\subsection{F2BP pre-impact relaxed and post-impact non-chaotic scenarios}
\begin{figure}[htbp]
	\centering
	\includegraphics[width=0.6\textwidth]{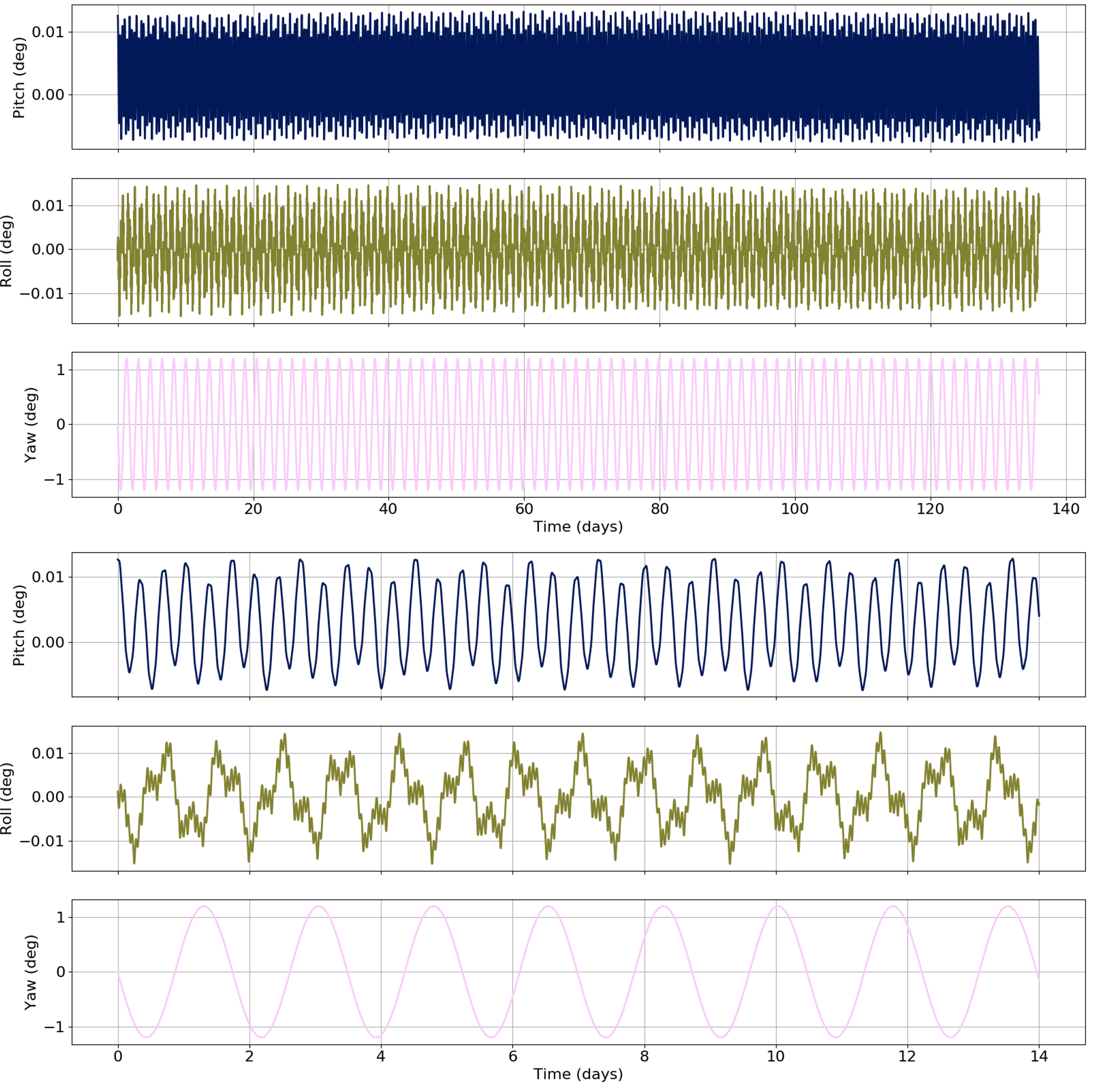}
	\caption{Pre-impact F2BP propagated attitude of Dimorphos (Euler 3-2-1 angles). The bottom panel shows the same results for a shorter time window, highlighting the short time-scale frequencies.}
	\label{figdimoeulerpre}
\end{figure}

Herein, we report the pre-impact relaxed and post-impact non-chaotic scenarios presented in Gramigna et al. (2024) \cite{gramigna2024hera}. These two scenarios have been studied using a global ephemerides approach. They are reported here to provide context for the comparisons with the analyzed NPA scenarios we will present in the next sections. The pre-impact scenario is depicted in Figure \ref{figdimoeulerpre}. For Dimorphos, we observe roughly 1 degree of libration in yaw and 0.01 degrees in both pitch and roll. The orbit eccentricity is approximately 0.016, and the mean anomaly is approximately 0, indicating that Dimorphos is trapped at periapsis. This is not a fully relaxed case, as we did not fit the initial rotation conditions to the pre-impact lightcurve data. This accounts for the small librations in pitch and roll observed in Figure \ref{figdimoeulerpre}, as well as a free libration in yaw.

The post-impact case assumes that the system is in an excited non-chaotic state after the impact. The DART impact is introduced in the dynamical model as an instantaneous variation in Dimorphos's orbital velocity. The values for this velocity variation are obtained from the JPL solution \textit{Dimorphos\_s527}\cite{ssd_jpl}, resulting in $dv = [ 0.42\cdot 10^{-6}\ ,-2.80\cdot 10^{-6}\ ,0.42\cdot 10^{-6}\ ] \ \text{km/s}$ in the RTN components with respect to Didymos, where we introduced a normal component equal to the radial one.

\begin{figure}[htbp]
	\centering
	\includegraphics[width=0.6\textwidth]{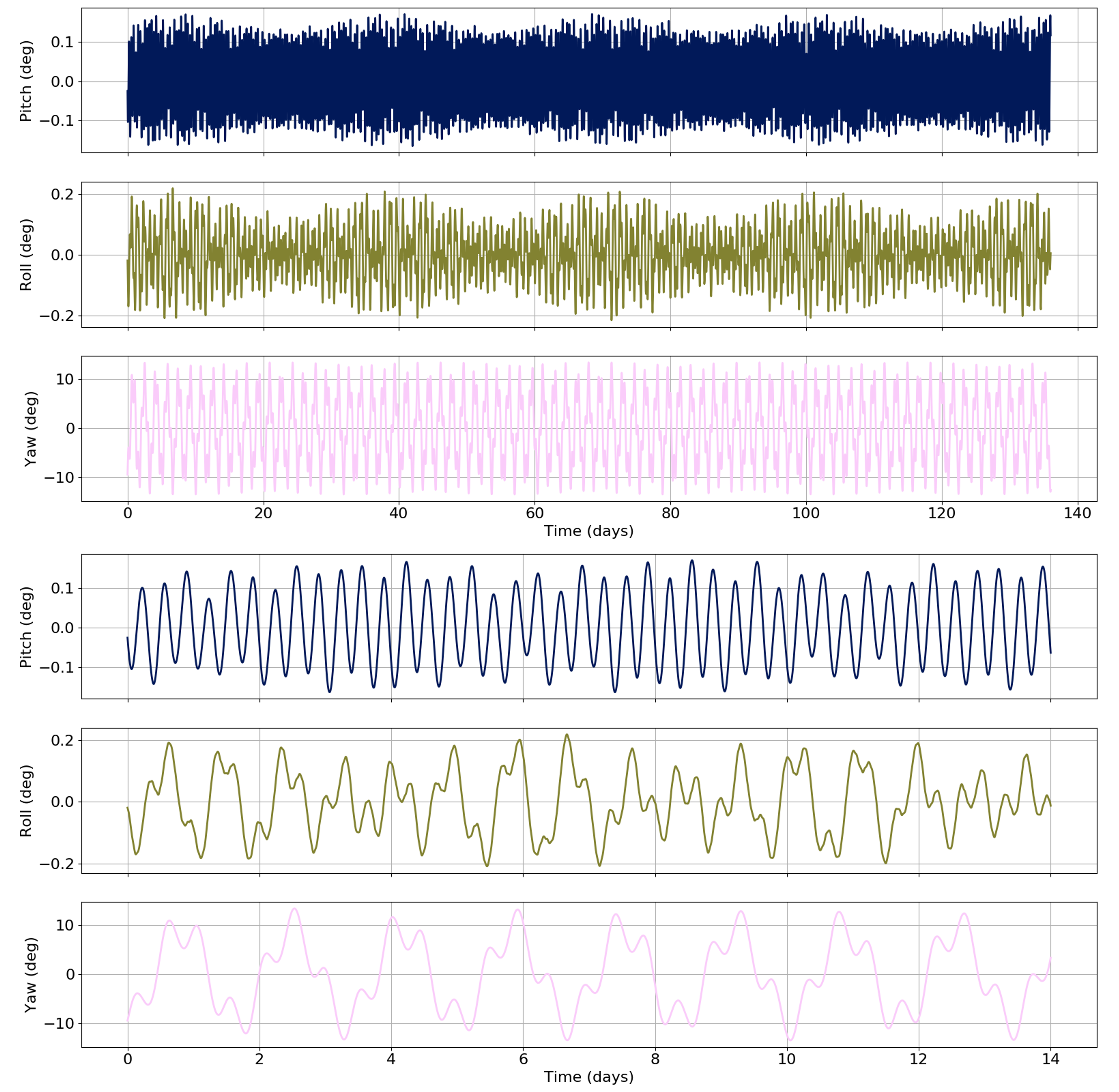}
	\caption{Post-impact F2BP propagated attitude of Dimorphos (Euler 3-2-1 angles). The bottom panel shows the same results for a shorter time window, highlighting the short time-scale frequencies.}
	\label{figdimoeulerpost}
\end{figure}

Figure \ref{figdimoeulerpost} presents the time variation of the resulting Euler angles for the post-impact case. Notably, we observe librations of roughly 10 degrees in yaw at multiple frequencies, while both pitch and roll exhibit librations of less than 1 degree at short (less than one day) and long (approximately one year) period frequencies. The eccentricity for this case is around 0.03, accompanied by a relative distance oscillation of roughly 80 meters. The mean anomaly maintains a clocking rate in line with the orbital rate. At the same time, the argument of periapsis undergoes a slow precession, consistent with the results in Meyer at al. (2023)\cite{meyer2023perturbed}. Dimorphos' pole nutation magnitude is approximately a factor of 10 higher than in the pre-impact case.

\subsection{F2BP post-impact non-principal axis scenarios}
As anticipated, the latest lightcurve observations indicate that Hera may find the Didymos system in a chaotic or NPA rotational state after DART's impact. 


Using F2BP simulations of the Didymos system, Pravec et al. (2024)\cite{pravec2024rotational} generated synthetic lightcurves in a Monte Carlo analysis of the post-DART dynamics. Comparing these synthetic lightcurves to seven detections of Dimorphos' lightcurves spanning from December 2022 through January 2023, they determined that Dimorphos is likely in a chaotic spin state. This finding is based on a statistical analysis of the evolution of Dimorphos' apparent magnitude. In this model, Dimorphos must be in a state of non-principal axis rotation to explain the observed changes in the lightcurve amplitude. They found that to obtain the best match with observations, the long axis of Dimorphos should be pointing, on average, toward or away from Didymos and chaotically tumbling around this synchronous or anti-synchronous state with tens of degrees amplitude in libration. Given this analysis, we must also consider scenarios where Hera finds Dimorphos in a chaotic NPA rotation state.

From the OD point of view, these scenarios are a more delicate and complex problem. Chaotic dynamical systems are characterized by a time limit beyond which it is impossible to forecast the system's state within a desired accuracy, also called the predictability horizon. Moreover, the analysis of Spoto et al. (2016) \cite{spoto2016shadowing} found a time limit called computability horizon, defined as the maximum number of observations that can be processed before the computation becomes numerically unstable. These \textit{horizons} are functions of the system's chaoticity and are related to the numerical instability of the orbit computation. Another work by Serra et al. (2018)\cite{serra2018multi} investigated numerical techniques to overcome the computability horizon limitation.

If Hera finds the Didymos system in a chaotic state, as predicted by Pravec et al. (2024)\cite{pravec2024rotational}, several of these aspects should be thoroughly investigated. These aspects include, for example, the selection of suitable durations for the gravity science arcs, propagators' tolerances, the evaluation of predictability and computability horizons, and the time-scale of chaos through the Lyapunov characteristic exponents (LCEs) (see Appendix D of Reference \cite{pravec2024rotational}). Unlike relaxed or non-chaotic cases, one of the main challenges with chaotic systems is obtaining a coherent global estimation of Dimorphos' relative orbit and rotational state for the entire mission duration. Such scenarios will likely face convergence issues, and the computation of partial derivatives may be prone to numerical and computational errors. Therefore, a global estimation approach may not be suitable for achieving a satisfactory OD fit with a full F2BP model and may result in unrealistic formal uncertainties. In this work, we propose a local approach for estimating asteroid ephemeris and rotational states in chaotic scenarios, which is a potential solution for accurately fitting the data. In general, in the multi-arc approach \cite{gramigna2024hera}, the solve-for parameters that form the state vector can be classified as: 
\begin{itemize}
    \item Global: parameters that do not vary in time and affect all the arcs simultaneously. They may include, among others, the GM of a celestial body, the spherical harmonics coefficients of its gravitational potential, bodies' moments of inertia, ephemeris and rotational state if exploiting a global approach;
    \item Local: parameters that only affect and depend on a single specific arc. Theoretically, any parameter can be treated as a local parameter. In the multi-arc approach, the S/C trajectory along each arc and the parameters related to its dynamics (e.g., Solar Radiation Pressure (SRP) scale factors, stochastic accelerations, and S/C attitude) are considered local parameters.
\end{itemize}

For studying the NPA cases, the bodies' position, velocities, Euler angles, and angular rates are treated as local parameters. Therefore, for each fly-by, the estimation considers a distinct state for the spacecraft and diverse states for Didymos and Dimorphos. After the estimation process, new "local" ephemerides for Didymos and Dimorphos are generated for each radio science arc, integrating from the newly determined states over a time interval corresponding to the arc duration. This approach does not ensure strict coherence in the satellite trajectory and rotation. However, like the classic multi-arc analysis for spacecraft trajectories, deterministic coherence gives way to stochastic coherence. In other words, the two states are deemed statistically compatible if the differences in adjacent orbits and frames fall within the respective uncertainty. When estimating Didymos' gravity field, the local approach is considered entirely equivalent to the global one. This equivalence is accompanied by anticipated higher uncertainties stemming from the absence of a uniquely coherent orbit for the entire mission and an increased number of estimated parameters. For this analysis, the adopted gravity science arcs have a duration of about 3-4 days.

\begin{figure}[h!]
	\centering
	\includegraphics[width=0.8\textwidth]{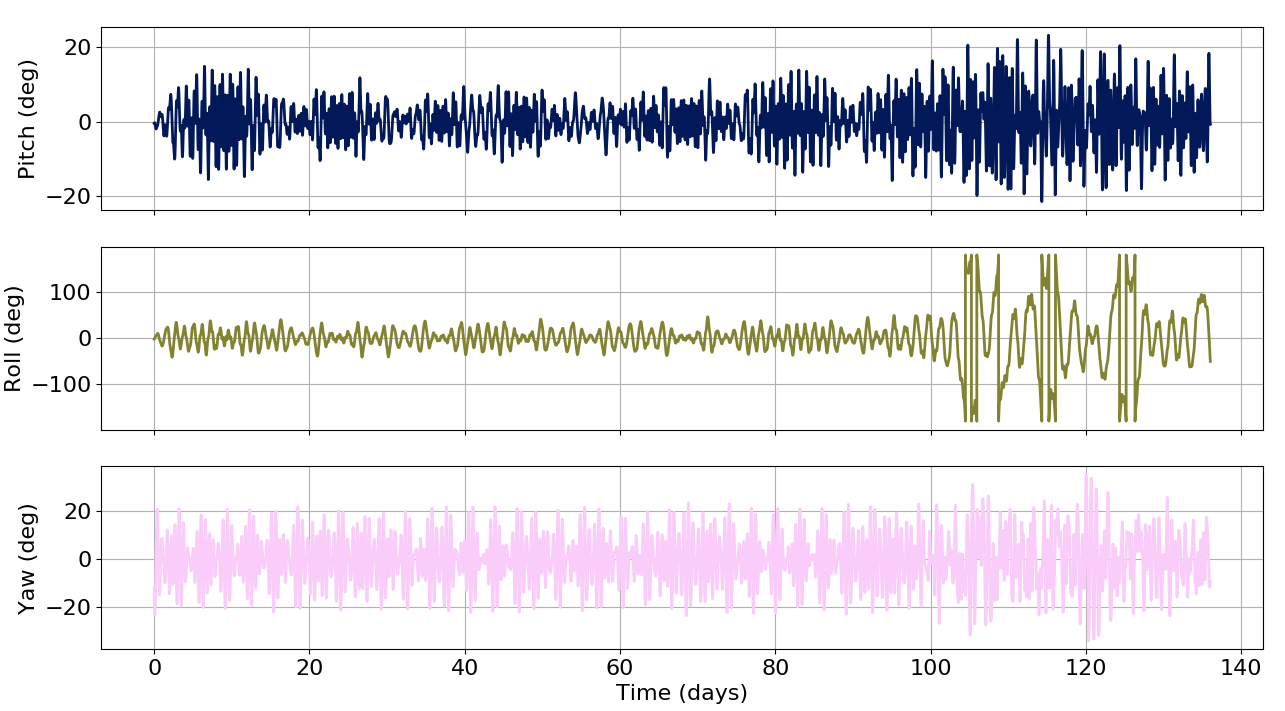}
	\caption{Post-impact F2BP propagated attitude of Dimorphos (Euler 3-2-1 angles) for the NPA case-1.}
	\label{figdimoeulerNPA1}
\end{figure}

\begin{figure}[h!]
	\centering
	\includegraphics[width=0.8\textwidth]{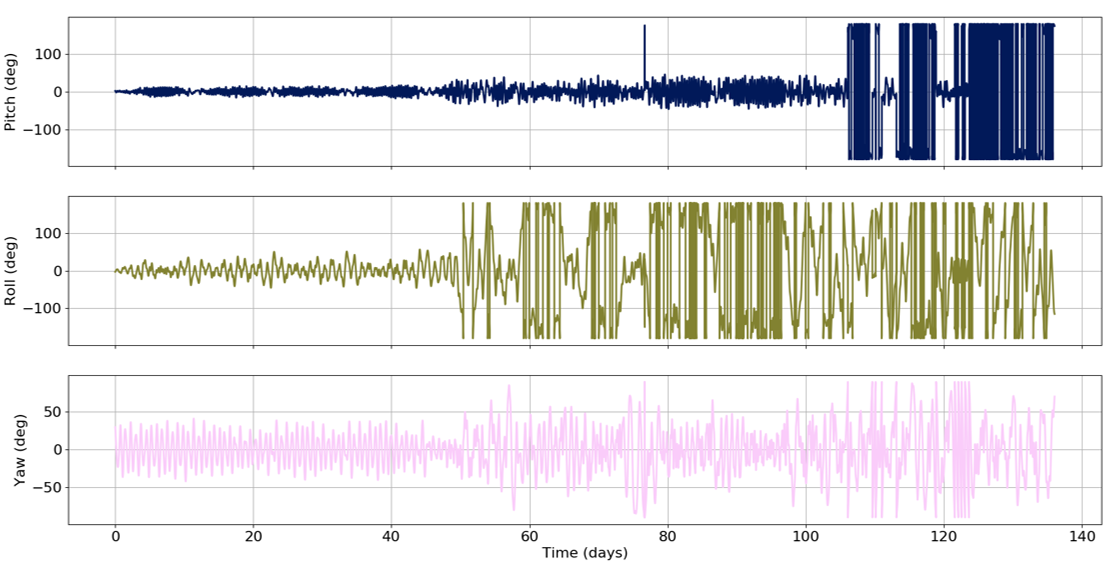}
	\caption{Post-impact F2BP propagated attitude of Dimorphos (Euler 3-2-1 angles) for the NPA case-2.}
	\label{figdimoeulerNPA2}
\end{figure}

\begin{figure}[h!]
	\centering
	\includegraphics[width=0.6\textwidth]{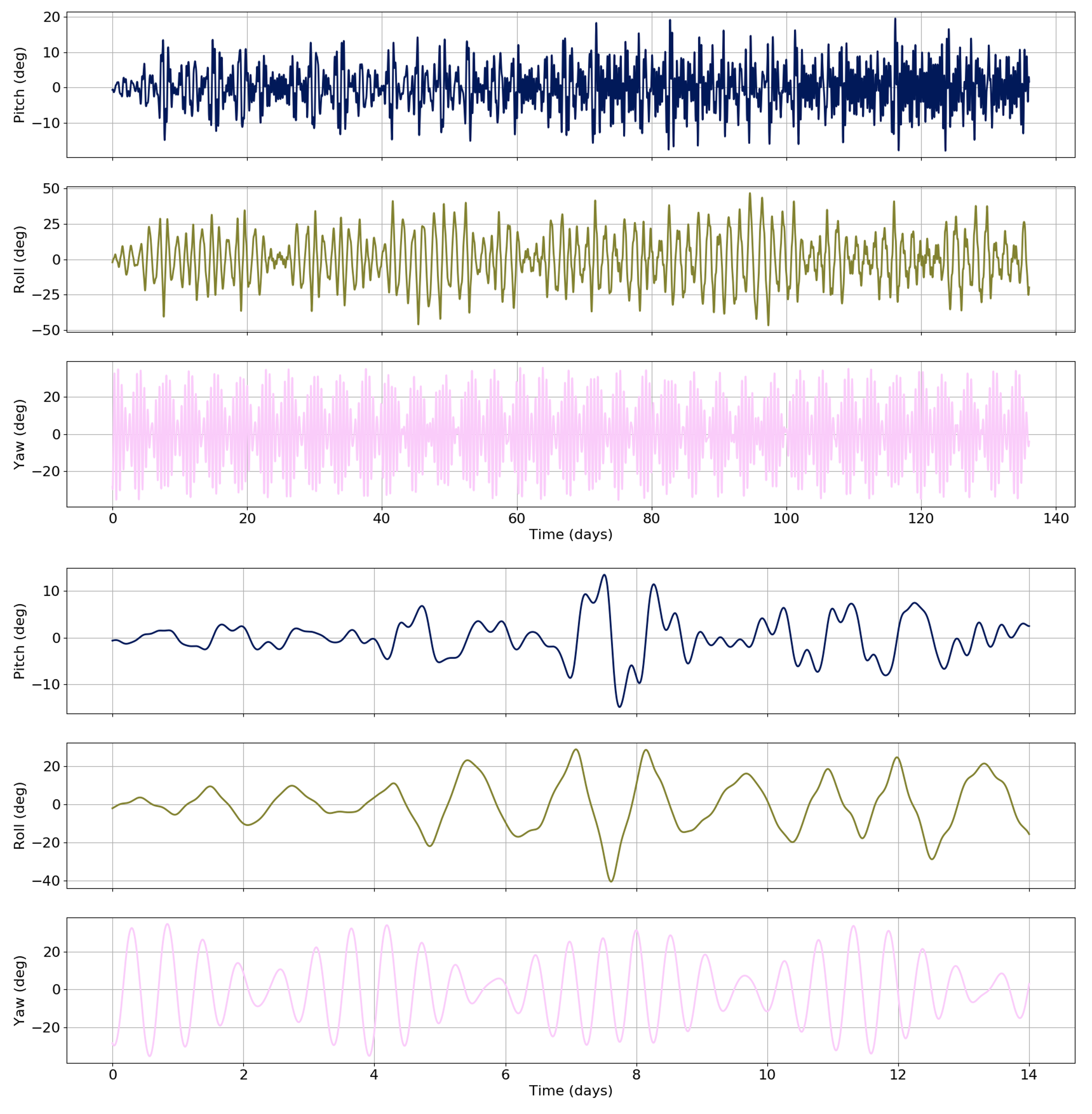}
	\caption{Post-impact F2BP propagated attitude of Dimorphos (Euler 3-2-1 angles) for the NPA case-3. The bottom panel shows the same results for a shorter time window, highlighting the short time-scale frequencies.}
	\label{figdimoeulerNPA3}
\end{figure}

\begin{figure}[h!]
	\centering
	\includegraphics[width=0.8\textwidth]{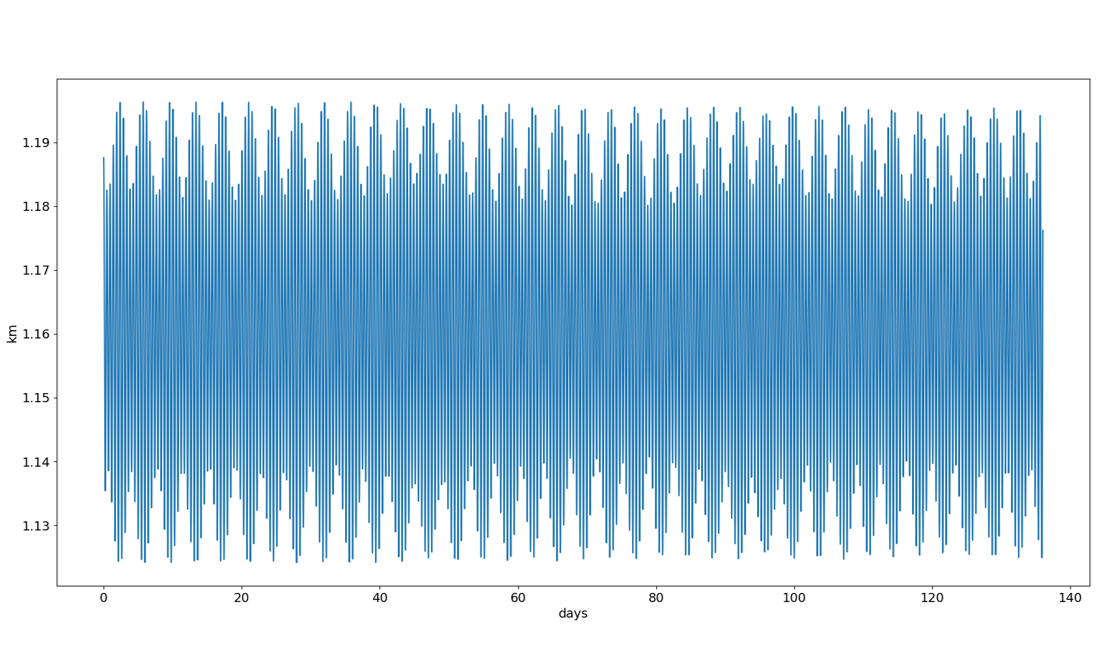}
	\caption{Post-impact relative distance between Didymos and Dimorphos (NPA case-3).}
	\label{figreldistNPA3}
\end{figure}

We analyzed three scenarios and performed covariance analyses among the NPA cases found by Pravec et al. (2024)\cite{pravec2024rotational} to match the light curves best. The simulated Euler angles for the cases are shown in Figures \ref{figdimoeulerNPA1}-\ref{figdimoeulerNPA2}-\ref{figdimoeulerNPA3}-\ref{figreldistNPA3}. Cases 1 and 2 correspond to the ones reported in Pravec et al. (2024)\cite{pravec2024rotational} as examples among the 16 best matches. In particular, case 1 has the largest probability value for both the mutual event offsets and cross-section variation. As can be noticed, both cases 1 and 2 show barrel instability, appearing much sooner in the second case. For case 1, the secondary is in NPA rotation with a synchronous configuration ($|$pitch$|$ and $|$yaw$|<90 $ deg). Case 2 shows the secondary having a NPA rotation around the anti-synchronous configuration ($|$yaw$|>90 $ deg) and switching between the synchronous and anti-synchronous states several times throughout the simulation. Finally, case 3 is similar to case 1 but without barrel instability.


\section{Results}

\begin{figure}[htbp]
	\centering
	\includegraphics[width=0.8\textwidth]{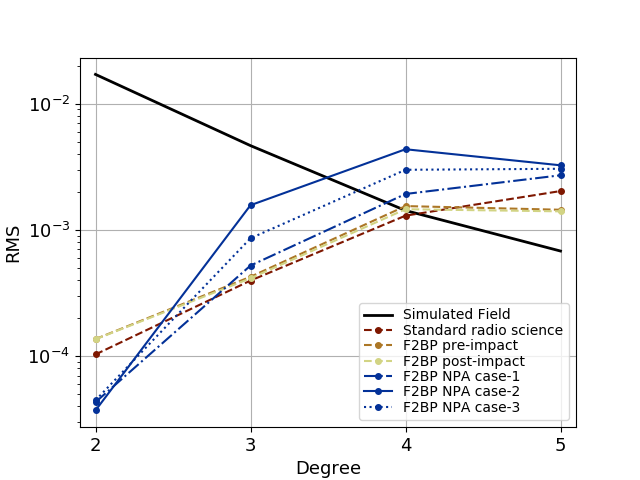}
	\caption{Power spectra of the extended gravity field of Didymos. Continuous line: simulated field; dashed lines: formal uncertainty of the estimated field at the end of the nominal mission under different scenarios. Red: standard radio science from Gramigna et al. (2024)\cite{gramigna2024hera}; light brown: F2BP pre-impact relaxed case from Gramigna et al. (2024)\cite{gramigna2024hera}; yellow: F2BP post-impact non-chaotic case from Gramigna et al. (2024)\cite{gramigna2024hera}; blue: F2BP NPA scenarios, plotted with different linestyles.}
	\label{figplotgravitynpa}
\end{figure}

\begin{figure}[h!]
	\centering
	\includegraphics[width=0.8\textwidth]{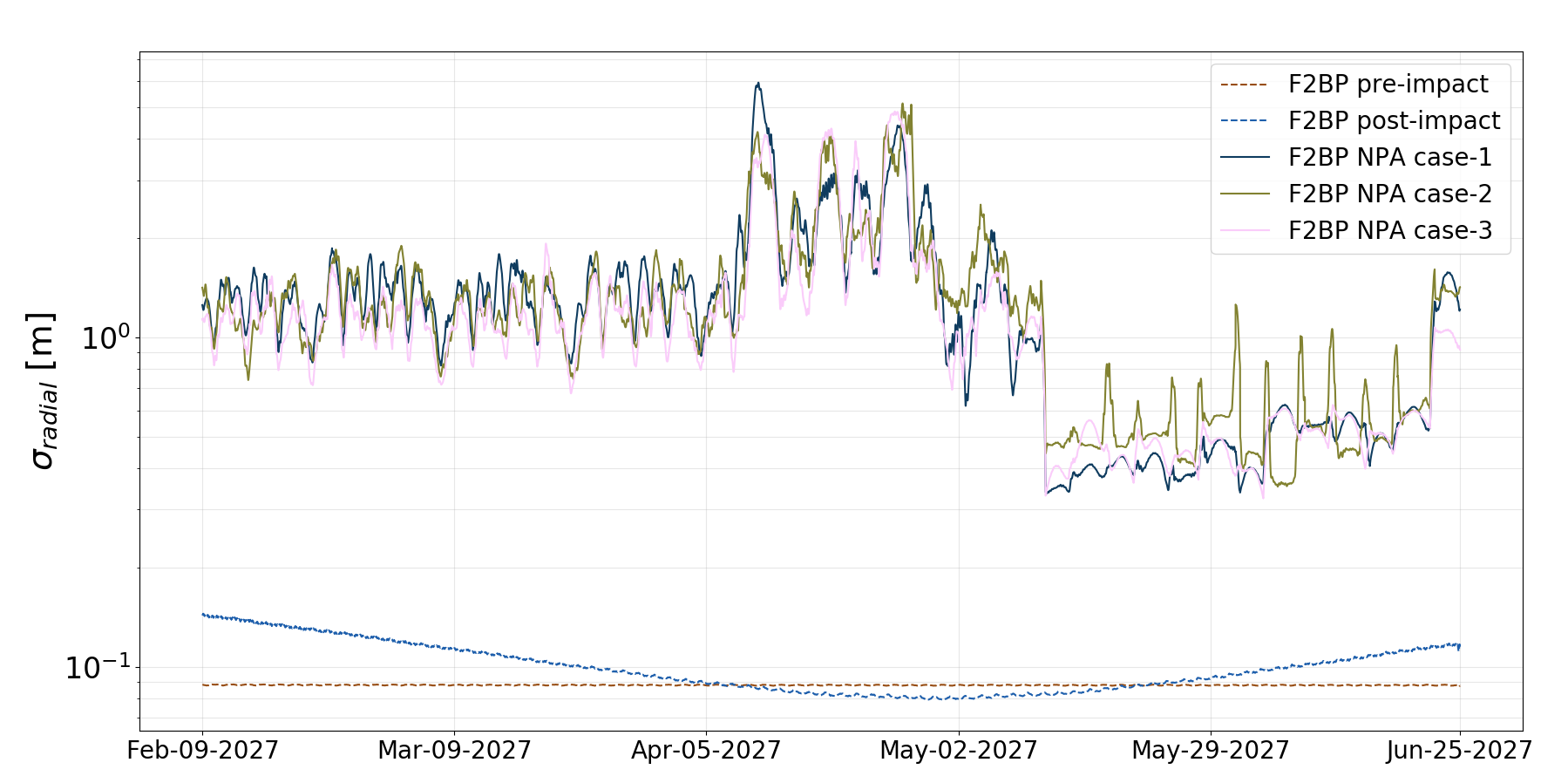}
	\caption{Expected 1$\sigma$ uncertainty of Dimorphos' position in the radial direction with respect to Didymos. The transverse and normal components show similar trend and are not reported here for brevity.}
	\label{figplotR}
\end{figure}

\begin{figure}[h!]
	\centering
	\includegraphics[width=0.8\textwidth]{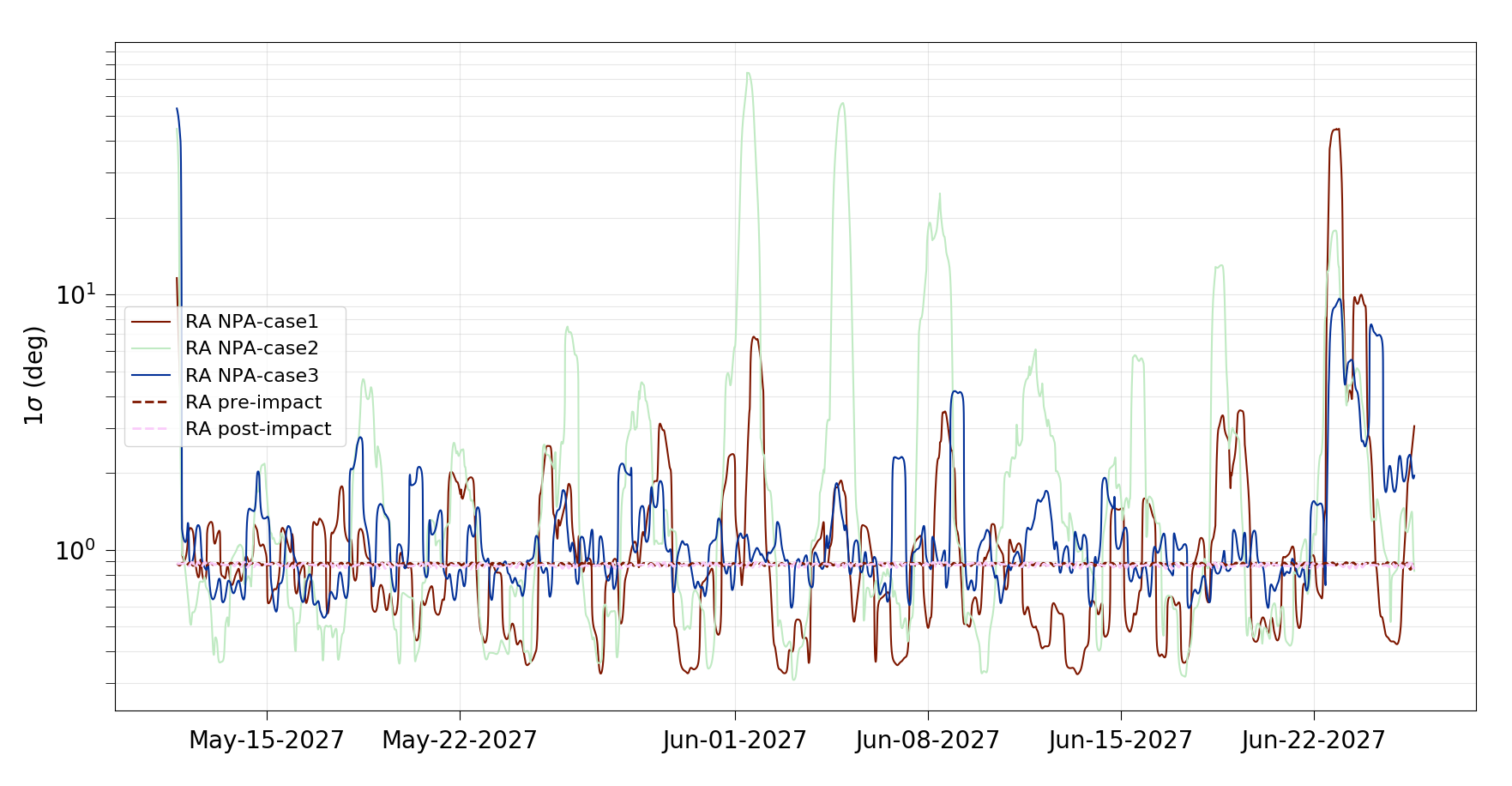}
	\caption{Expected 1$\sigma$ uncertainty of Dimorphos' right ascension frame parameter.}
	\label{figplotRaDimo}
\end{figure}

\begin{figure}[h!]
	\centering
	\includegraphics[width=0.8\textwidth]{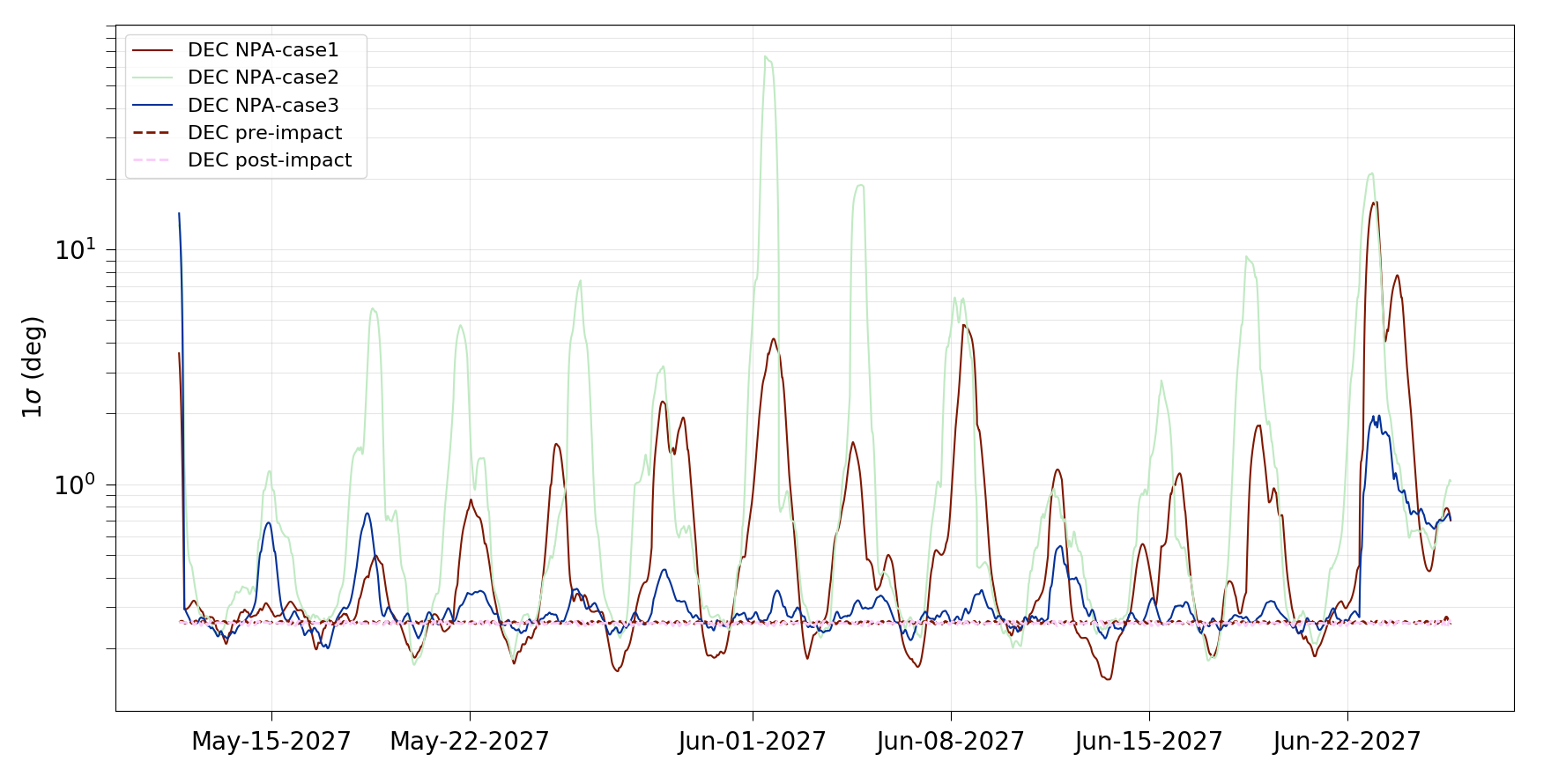}
	\caption{Expected 1$\sigma$ uncertainty of Dimorphos' declination frame parameter.}
	\label{figplotDecDimo}
\end{figure}

\begin{figure}[h!]
	\centering
	\includegraphics[width=0.8\textwidth]{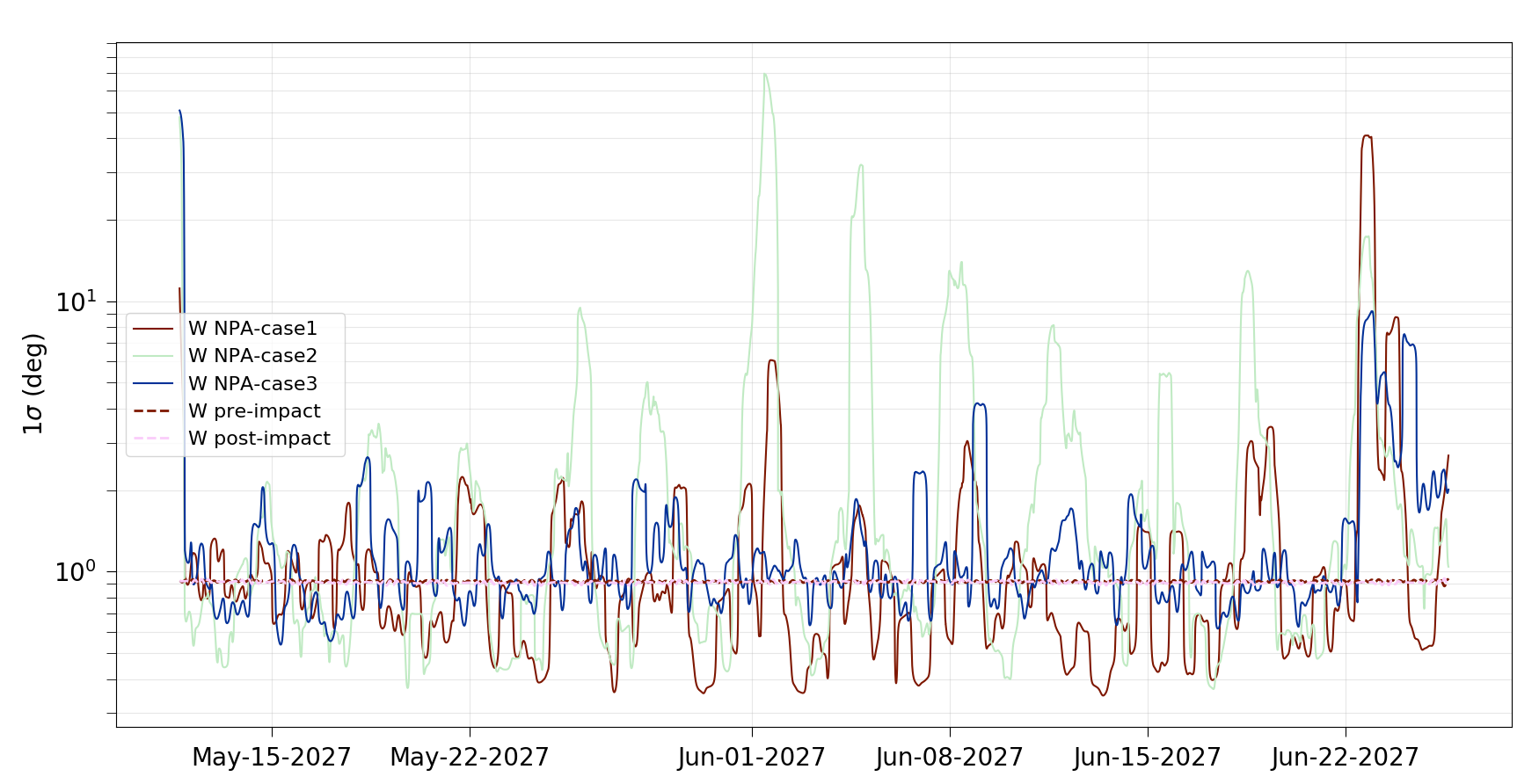}
	\caption{Expected 1$\sigma$ uncertainty of Dimorphos' prime meridian frame parameter.}
	\label{figplotWDimo}
\end{figure}

Our previous work \cite{gramigna2024hera} demonstrates that the pre-impact relaxed scenario and the non-chaotic post-impact scenario show comparable uncertainties in the main scientific parameters of interest to the ones retrieved adopting standard radio science models. This behavior was expected, as the orbits of the two analyses are similar and exhibit a comparable translational motion, showcasing non-uniform (or non-sinusoidal) orbit deviations induced by the F2BP interaction. In this context, the F2BP allows for a further improvement in Dimorphos' orbit estimation thanks to the coupled motion, which helps lower the correlations between Dimorphos' translational and rotational states. 
Following that preliminary investigation, the Hera RSE covariance analysis results for the three of NPA scenarios described above are reported in Tables \ref{tabpaperold}-\ref{tabgmj2npa}-\ref{tabmomentsofinertia} and Figures \ref{figplotgravitynpa}-\ref{figplotR}-\ref{figplotRaDimo}-\ref{figplotDecDimo}-\ref{figplotWDimo}. 

First, Table \ref{tabpaperold} provides the estimated GMs and $J_2$ coefficients from our previous work \cite{gramigna2024hera}, while Table \ref{tabgmj2npa} reports the 1$\sigma$ uncertainties for the newly analyzed NPA cases. We can observe a factor 1.3-3 degradation in the bodies GMs and $\sim$ 2 in Didymos' $J_2$. This result was expected and related to the local estimation of the bodies' dynamical parameters. However, the degradation for the NPA cases is not dramatic, and the results still satisfy the Hera RSE requirements. Interestingly, thanks to the NPA scenarios, Dimorphos' $J_2$ uncertainty improves by a factor $\sim$ 2.2 compared to the previous F2BP cases. Comparing the three NPA cases, we observe that case-2, which is the most chaotic one, shows a factor $\sim$ 2 degradation in the asteroids' GMs uncertainty with respect to case-1 and -3.

Furthermore, Figure \ref{figplotgravitynpa} illustrates the uncertainty in the recovered gravity field of Didymos at the end of Hera's nominal mission under different simulation scenarios. The plot provides a quick understanding of the expected performance in retrieving the spherical harmonic coefficients. The black curve represents the magnitude of the simulated gravity field for each degree. At the same time, the colored lines indicate the accuracy of each test case, measured as the RMS of the estimated formal covariances at a given degree. The corresponding curve must lie below the simulated field curve to observe a specific degree of Didymos' gravity field. In Figure \ref{figplotgravitynpa} we also reported the "Standard radio science", "F2BP pre-impact", and "F2BP post-impact" scenarios from our previous investigation \cite{gramigna2024hera} to provide context. The three NPA scenarios allow for a better recovery of Didymos degree 2. At the same time, they show higher uncertainties for Didymos' higher degrees (particularly for case-2, which is the most chaotic one). In addition, the improvement in degree 2 is about a factor of 2-2.5. This effect is linked to better estimating the bodies' moments of inertia (MOI) thanks to higher observability of the system's dynamics when Dimorphos is in NPA state, see Table \ref{tabmomentsofinertia}. Overall, the three analyzed NPA cases show similar results in the MOI uncertainties, resulting in more than one order of magnitude improvement compared to the F2BP post-impact non-chaotic scenario studied in our previous work. Indeed, the un-normalized second-degree gravity coefficients are related to the moments of inertia of the bodies \cite{scheeres2012orbital}, hence a higher observability and sensibility to the asteroids MOI reflects in a better knowledge of their degree 2 as well. 

\renewcommand{\arraystretch}{1.3} 
\begin{table}[htbp]
   \fontsize{10}{10}\selectfont
    \caption{Estimated formal $1\sigma$ uncertainties of Didymos and Dimorphos GM and $J_2$ from Gramigna et al. (2024)\cite{gramigna2024hera}.}
   \label{tabpaperold}
        \centering
   \begin{tabular}{|l|l|l|l| } 
      \hline 
	\textbf{Coefficient} & \textbf{Standard} & \textbf{F2BP} & \textbf{F2BP} \\
    &\textbf{radio science}& \textbf{pre-impact} & \textbf{post-impact} \\
    & & \textbf{relaxed} & \textbf{non-chaotic} \\
      \hline 
	\textbf{Didymos} & & &  \\
	GM (km$^3$/s$^2$) & 1.61$\cdot10^{-12}$ & 1.54$\cdot10^{-12}$ & 1.45$\cdot10^{-12}$ \\
	$J_2$ & 7.02$\cdot10^{-5}$ & 2.32$\cdot10^{-5}$ & 2.16$\cdot10^{-5}$ \\	
	\hline
	\textbf{Dimorphos} & & &  \\
	GM (km$^3$/s$^2$) & 2.73$\cdot10^{-13}$ & 3.48$\cdot10^{-13}$ & 3.35$\cdot10^{-13}$ \\
	$J_2$ & 5.44$\cdot10^{-3}$ & 5.64$\cdot10^{-3}$ & 4.21$\cdot10^{-3}$ \\
      \hline
   \end{tabular}
\end{table}

\renewcommand{\arraystretch}{1.3} 
\begin{table}[htbp]
	\fontsize{10}{10}\selectfont
    \caption{Estimated formal $1\sigma$ uncertainties of Didymos and Dimorphos GM and $J_2$ for the three NPA scenarios studied in this work.}
   \label{tabgmj2npa}
        \centering
   \begin{tabular}{|l|l|l|l| } 
      \hline 
	\textbf{Coefficient} & \textbf{F2BP NPA} & \textbf{F2BP NPA} & \textbf{F2BP NPA} \\
    & \textbf{case-1} & \textbf{case-2}  & \textbf{case-3}\\
    \hline 
	\textbf{Didymos} & & & \\
	GM (km$^3$/s$^2$) & 2.15$\cdot10^{-12}$ & 4.12$\cdot10^{-12}$ & 2.77$\cdot10^{-12}$\\
	$J_2$ & 4.21$\cdot10^{-5}$ & 4.87$\cdot10^{-5}$ & 4.56$\cdot10^{-5}$\\	
	\hline
	\textbf{Dimorphos} & & & \\
	GM (km$^3$/s$^2$) & 5.10$\cdot10^{-13}$ & 1.09$\cdot10^{-12}$  & 6.71$\cdot10^{-13}$ \\
	$J_2$ & 2.14$\cdot10^{-3}$ & 1.93$\cdot10^{-3}$ & 1.15$\cdot10^{-3}$ \\
      \hline
   \end{tabular}
\end{table}

\begin{table}[htbp]
	\fontsize{10}{10}\selectfont
    \caption{Estimated formal uncertainties ($1\sigma$) in the moments of inertia given by the F2BP scenarios of Gramigna et al. (2024)\cite{gramigna2024hera} and the three NPA scenarios studied in this work. The values are rounded to 2 decimal places and provided in kg$\cdot$km$^2$, while the apriori covariances are unconstrained.}
   \label{tabmomentsofinertia}
        \centering
   \begin{tabular}{|l|l|l|l|l|l| } 
      \hline 
	\textbf{Coefficient} & \textbf{F2BP} & \textbf{F2BP} & \textbf{F2BP NPA} & \textbf{F2BP NPA} & \textbf{F2BP NPA} \\
    &\textbf{pre-impact} & \textbf{post-impact} & \textbf{case-1}& \textbf{case-2} & \textbf{case-3}\\
    & \textbf{relaxed} & \textbf{non-chaotic} & & &\\
      \hline 
	\textbf{Didymos} & & & & & \\
	$I_{XX}$ & 5.46$\cdot10^{10}$  & 5.19$\cdot10^{9}$ & 2.31$\cdot10^{8}$ & 3.70$\cdot10^{8}$ & 2.42$\cdot10^{8}$\\
	$I_{YY}$ & 5.46$\cdot10^{10}$  & 5.19$\cdot10^{9}$ & 2.31$\cdot10^{8}$ & 3.70$\cdot10^{8}$ & 2.42$\cdot10^{8}$ \\
    $I_{ZZ}$ & 5.46$\cdot10^{10}$  & 5.19$\cdot10^{9}$ & 2.31$\cdot10^{8}$ & 3.70$\cdot10^{8}$ & 2.42$\cdot10^{8}$\\
	$I_{XY}$ & 1.42$\cdot10^{7}$  & 1.40$\cdot10^{7}$ & 3.97$\cdot10^{6}$ & 2.91$\cdot10^{6}$ & 3.49$\cdot10^{6}$ \\  
    $I_{XZ}$ & 4.08$\cdot10^{7}$  & 4.08$\cdot10^{7}$ & 9.26$\cdot10^{6}$ & 5.57$\cdot10^{6}$ & 9.09$\cdot10^{6}$\\
	$I_{YZ}$ & 3.35$\cdot10^{6}$  & 3.36$\cdot10^{6}$ & 3.72$\cdot10^{6}$ & 2.25$\cdot10^{6}$ & 3.66$\cdot10^{6}$\\
	\hline
	\textbf{Dimorphos} & & & & & \\
	$I_{XX}$ & 7.78$\cdot10^{5}$  & 5.36$\cdot10^{5}$ & 2.81$\cdot10^{5}$ & 1.29$\cdot10^{5}$ & 1.40$\cdot10^{5}$\\
	$I_{YY}$ & 8.08$\cdot10^{5}$  & 5.57$\cdot10^{5}$ & 3.76$\cdot10^{5}$ & 1.61$\cdot10^{5}$ & 2.03$\cdot10^{5}$\\
    $I_{ZZ}$ & 1.06$\cdot10^{6}$  & 7.64$\cdot10^{5}$ & 4.33$\cdot10^{5}$ & 2.02$\cdot10^{5}$ & 2.21$\cdot10^{5}$\\
	$I_{XY}$ & 3.28$\cdot10^{3}$  & 3.12$\cdot10^{3}$ & 2.08$\cdot10^{4}$ & 9.77$\cdot10^{3}$ & 2.20$\cdot10^{4}$\\  
    $I_{XZ}$ & 2.70$\cdot10^{4}$  & 2.70$\cdot10^{4}$ & 3.05$\cdot10^{4}$ & 1.99$\cdot10^{4}$ & 2.66$\cdot10^{4}$\\
	$I_{YZ}$ & 1.16$\cdot10^{4}$  & 1.12$\cdot10^{4}$ & 4.63$\cdot10^{3}$ & 4.58$\cdot10^{3}$ & 2.64$\cdot10^{3}$\\
      \hline
   \end{tabular}
\end{table}

Figure \ref{figplotR} shows the 1$\sigma$ uncertainties in Dimorphos' relative orbit, given in RTN coordinates with respect to Didymos, and averaged over one orbital period to improve readability. For brevity, we report only the radial component, as the transverse and normal components exhibit similar trends. Again, the "pre-impact" and "post-impact" curves are from our previous work and show higher accuracy due to their global ephemerides estimation approach. Adopting the local dynamical state estimation, on average, degrades the reconstructed orbit by about one order of magnitude. Therefore, in the next steps of this research, it will be crucial to implement a constrained multi-arc orbit determination setup, which is expected to improve the performance of the local approach while still retaining its computational efficiency.

Finally, Figures \ref{figplotRaDimo}-\ref{figplotDecDimo}-\ref{figplotWDimo} present the formal uncertainties in Dimorphos' rotational state. Specifically, we provide the 1-$\sigma$ uncertainties for the body-fixed right ascension, declination, and prime meridian relative to the inertial frame, averaged over one rotational period. Overall, the NPA cases exhibit larger variations in the frame parameters due to the chaotic spin state compared to the non-chaotic scenarios, although the average values remain comparable. Additionally, as expected, NPA case-3, being the least chaotic, shows lower variations in frame parameters than NPA case-2, which is the most chaotic and has higher amplitude variations. As with the relative orbit, a constrained multi-arc orbit determination setup is anticipated to enhance the performance of the local approach in rotational state estimation.


\section{Conclusion and Future Work}
If Hera finds the Didymos system in a non-chaotic rotational state, the standard radio science models and the F2BP should provide comparable estimates of the main parameters of interest, particularly Dimorphos' mass, which is the primary objective of the Hera mission. The choice between adopting the F2BP or the standard approach will depend on selecting the model that best fits the data during the operational mission.

However, recent observations suggest that Hera will likely encounter the system in a chaotic or tumbling state, for which the F2BP dynamical model will be crucial for accurately fitting the data. Our work demonstrates that a local estimation approach is suitable to fit the asteroids' translational and rotational states in chaotic or tumbling cases, as opposed to the standard global approach used for non-chaotic scenarios. The NPA rotation scenarios will likely face convergence issues and higher computational demands for partial derivatives computation, which may result in numerical errors. Thus, a global estimation may not be suitable for achieving a satisfactory OD fit with an F2BP model and may yield unrealistic formal uncertainties. In this context, the local approach to estimation emerges as an effective solution.

As expected, the presented F2BP covariance analysis highlighted slightly lower accuracies in the solved-for dynamical parameters due to the local estimation approach. The degradation factor in Didymos' and Dimorphos' gravitational parameters is between 1.3 and 3. Regarding Dimorphos' relative orbit, we observed a degradation of about one order of magnitude (on average) compared to the estimation of global ephemerides. Nevertheless, the recovery of the bodies' moments of inertia improved by more than one order of magnitude compared to the F2BP post-impact non-chaotic scenario explored in our previous work. Consequently, the degree-2 gravity field improved by a factor of 2.5 in the NPA rotation cases. This improvement is attributed to the higher observability of the system's dynamic state when in a chaotic and tumbling regime. Regarding Dimorphos' rotational state, the formal uncertainties in the frame parameters for the NPA cases exhibit larger variations compared to the non-chaotic cases, while the average values remain comparable.

The next step in this research is to develop a constrained multi-arc orbit determination setup. Imposing constraints on the dynamical state of the asteroids and the spacecraft at the boundaries of the OD arcs, this setup allows for greater coherency in the orbit reconstruction, improving the filter's convergence and performance. This approach should minimize the differences between the global estimation used in standard radio science experiments and the local estimation described in this work, representing a more conservative scenario. Additional analyses shall also be performed to select suitable durations for the gravity science arcs (e.g., depending on the Lyapunov time and characteristic exponents) and propagator tolerances. Furthermore, we should elaborate on the concepts of predictability and computability horizons as a function of the Hera mission's phases.

\section{Acknowledgment}
EG, RLM, MZ, and PT wish to acknowledge Caltech and the NASA Jet Propulsion Laboratory for granting the University of Bologna a license to an executable version of MONTE Project Edition S/W. EG, RLM, MZ, PT, and GT are grateful to the Italian Space Agency (ASI) for financial support through Agreement No. 2022-8-HH.0 in the context of ESA's Hera mission.


\bibliographystyle{AAS_publication}   
\bibliography{references}   

\end{document}